\documentclass[11pt]{cernrep}
 \usepackage{epsfig}
  \usepackage{amsmath}
   \usepackage{amssymb}
    \usepackage{pifont}
     \usepackage{here}
\begin{document}
\title{%
\hspace*{0.8\textwidth}\begin{minipage}{0.2\textwidth}
\vspace*{-10mm}\textsc{\small RUB-TPII-16/03}\end{minipage}\\
 WHAT IS THIS THING CALLED PION DISTRIBUTION AMPLITUDE? FROM THEORY TO DATA}
\author{Alexander~P.~Bakulev$^a$, S.~V.~Mikhailov$^a$, N.~G.~Stefanis$^b$}
\institute{$^a$~Bogoliubov Laboratory of Theoretical Physics,
           JINR, 141980, Moscow Region, Dubna, Russia\\
           $^b$~Institut f\"ur Theoretische Physik II,
           Ruhr-Universit\"at Bochum, D-44780 Bochum, Germany}
\maketitle
\begin{abstract}
We discuss the status of the pion distribution amplitude (DA)
from  analyzing the CLEO experimental data in the context
of QCD sum-rule techniques and QCD perturbation theory at the NLO accuracy.
The constraints extracted this way for the Gegenbauer coefficients
$a_2$ and $a_4$
exclude $\Phi_\text{CZ}$ at the $4\sigma$ level,
while $\Phi_\text{asy}$ is outside the $3\sigma$ error ellipse.
These data provide strong support for the type of endpoint-suppressed,
double-humped pion DA we derived via QCD sum rules with nonlocal condensates
and favor a value of the vacuum quark virtuality
$\lambda_q^2\simeq 0.4$~GeV${}^2$.
This pion DA is in agreement with the E791 data,
though these experimental results should be viewed carefully
and further confirmation is necessary for a more accurate
judging of pion DAs from them.
\end{abstract}

\section{INTRODUCTION}

It is widely believed today that the nontrivial QCD vacuum plays
an important role in understanding the analytic properties
of hadron distribution amplitudes (DA)
in terms of their quark and gluon degrees of freedom \cite{CZ84,Ste99}.
In fact, one can use QCD sum rules with nonlocal condensates \cite{MR86}
to connect dynamic properties of (light) mesons,
like form factors and DAs, directly with the QCD vacuum.
The classical example is the pion DA,
which describes how the pion's longitudinal momentum
is shared between its quark and antiquark constituents
when probed at large momentum transfer $Q^2$.
First, a detailed knowledge of the pion DA is necessary
in order to make precise calculations of ``hard-scattering'' form factors,
like $F_{\pi}^{\text{em}}\left(Q^{2}\right)$
and $F_{\pi\gamma}\left(Q^{2}\right)$
and compare the results with the experimental data.
Second, the parton structure is interesting
in its own right, providing insight
into the nonperturbative hadron structure at large distances.
There is a long history of determining the pion DA
starting from the asymptotic limit of perturbative QCD \cite{ER80,LB80}
to QCD sum rules \cite{CZ84,MR86,BF89,BJ97,BM98,BMS01},
to instanton-based models \cite{PPRWK99,PR01,ADT00},
and to lattice computations \cite{Lattice}.
Now, with the help of the recent CLEO data \cite{CLEO98},
one gets a handle on the pion DA from the experimental side.
Indeed, these data can be analyzed \cite{SchmYa99,BMS02}
using the framework of light cone sum rules \cite{Kho99}
to extract constraints on the Gegenbauer coefficients $a_2$ and $a_4$
of the conformal expansion.
Supplementary constraints on the shape of the pion DA
are supplied by diffractive di-jets production events \cite{E79102},
but the uncertainties of these experimental results
and their controversial theoretical interpretations \cite{Che01,NSS01,BISS02}
are still prohibiting definite conclusions.

\section{NONLOCAL CONDENSATES: THE MODUS OPERANDI}
\subsection{Modelling the nonperturbative QCD vacuum}
The nonlocal quark condensate represents a partial resummation
of the OPE to all orders in terms of the vacuum expectation value
of the nonlocal operator
\begin{eqnarray}
 \langle :\!\bar{q}_{\sigma}(0)E(0,z)q_{\rho}(z)\!:\rangle
  = \frac{\langle \bar{q}q \rangle}{4}
    \left[F_\text{S}(z^{2})-\frac{i\hat z}{4}F_\text{V}(z^2)
    \right]_{\rho \sigma},\quad \!\!
     E(0,z)
  = P\exp\!\left[-ig_s\!\!\int_0^z \!\!A_\mu(y)dy^\mu\right],
 \label{eq:NonQ}
\end{eqnarray}
where $\sigma,~\rho$ are spinor indices and the integral
in the Fock--Schwinger string $E(0,z)$
is taken along a straight-line path.
Note that $F_\text{S,V}(z^2)$ are analytic functions around the origin
and that their derivatives at zero are related
to condensates of corresponding dimension.
Recall that the condensates of lowest dimensions
\begin{equation}
 \label{eq:Q6}
 Q^3 = \langle \bar{q}q \rangle \,,\quad
 Q^5 =
   ig_s\langle \bar{q}{G_{\mu\nu}\sigma_{\mu\nu}}q \rangle
   \equiv m_0^2 \cdot Q^3\,,\quad
  Q^6 = \langle t^aJ^a_\mu t^bJ^b_\mu\rangle
\end{equation}
form the basis of the standard QCD sum rules \cite{SVZ79}
and have been estimated,
while higher-dimensional ones are yet unknown.
In the chiral limit one has
\begin{equation}
\label{eq:lambda_q}
  4\frac{d F_\text{S}(z^{2})}{d z^{2}}|_{z=0}
 = \frac{Q^5}{4Q^3}
 \equiv
   \frac{m^2_0}{4}
 = \frac{\lambda_q^2}2
\end{equation}
and the parameter $\lambda_q^2/2$ fixes the width of
$F_\text{S}(z^2)$ around the origin.

For not too large Euclidean distances $z^2=-z_\text{E}^2>0$, the
nonlocality behavior of the (quark) condensate can be implemented
by the Gaussian ansatz \cite{MR86}
$
 \label{eq:Gauss}
  F_\text{S}^\text{G}(z^2)
  = \exp\left(-\lambda_q^2z^2/8 \right),
$
in which $\lambda_q$ has the meaning
of an inverse vacuum quark correlation length.
This corresponds to the specific form of the virtuality distributions
$f_\text{S} = \delta\left(\alpha - \lambda_q^2/2\right)$,
$f_\text{V} \sim \alpha_s Q^3 \delta'\left(\alpha - \lambda_q^2/2\right), \ldots$
~\cite{MR86,BM98}
with
\begin{equation}
 \label{eq:fsv}
  F_\text{S,V}(z^2)
   = \int_{0}^{\infty} e^{-\alpha z^2/4}\,
      f_\text{S,V}(\alpha)\, d\alpha\,,
 \quad \text{where~}
 \int_{0}^{\infty}\!\!  f_\text{S,V}(\alpha)\,  d\alpha =
  \Big\{\vphantom{\Big\}}
   \begin{array}{ll}
       1,& \mbox{S-case};\\
       0, & \mbox{V-case,~chiral limit.}
             \end{array}
\end{equation}
This kind of virtuality distributions fix only \textit{one} main property
of the nonperturbative vacuum---quarks can flow through the vacuum
with a nonzero momentum $k$,
and the average virtuality of such vacuum quarks is just
$\langle k^2\rangle = \lambda_q^2/2$
(for a determination of $\lambda_q^2/2$ from lattice data,
see \cite{BM02}).

\subsection{Nonlocal QCD sum rules and the pion distribution amplitude}
\begin{figure}[hbt]
\centering{\epsfig{file=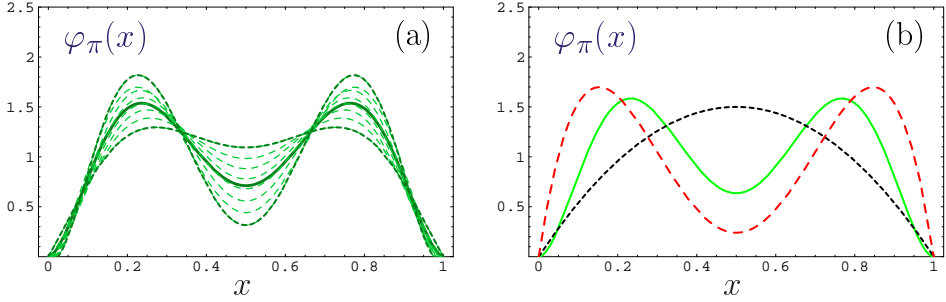,height=4cm,width=12.7cm,clip=}}
 \vspace{0.0cm}\caption{(a)
  Bunch of pion DAs for $\lambda_q^2=0.4$~GeV$^2$ determined
  with nonlocal QCD sum rules at normalization scale $\mu\approx 1.0$~GeV.
  (b) The BMS model \cite{BMS01} (solid line) in comparison
  with the asymptotic DA (dotted line)
  and the CZ model (dashed line) \cite{CZ84}.}
 \label{fig:bunch}
\end{figure}

The pion DA of twist 2, $\varphi_{\pi}(x,\mu^2)$
($x$ being here the longitudinal momentum fraction),
defined by
\begin{equation}
\label{eq:defpionda}
 \langle 0\mid \bar
d(z)\gamma^{\mu}\gamma_5 E(z,0) u(0)\mid \pi(P)\rangle \Big|_{z^2=0}
 = i f_{\pi}P^{\mu}
    \int^1_0 dx e^{ix(zP)}\ \varphi_{\pi}(x,\mu^2)\; ,
\end{equation}
can be related to the nonlocal condensates
by means of the following sum rule
\begin{equation}
\label{eq:nlcsrda}
  f_{\pi}^2\varphi_\pi(x)
=
  \int_{0}^{s_{\pi}^0}\rho^\text{pert}(x;s)e^{-s/M^2}ds
   + \frac{\langle \alpha_s GG\rangle}{\pi 24 M^2}\Phi_G(x;M^2)
   + \frac{16\pi\alpha_s\langle{\bar{q}q\rangle}^2}{81M^4}
   \sum_{i=S,V,T_j}\Phi_i(x;M^2) \; ,
\end{equation}
\begin{figure}[hbt]
 \centering{\epsfig{file=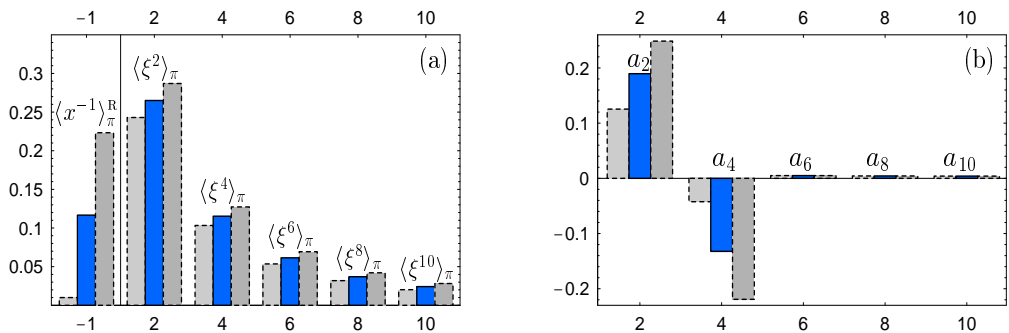,height=4.5cm,width=14.0cm,clip=}}
  \vspace{0.0cm}\caption{(a) First nonzero moments
   $\langle{\xi^N}\rangle_\pi$ (up to $N=10$, see text) and
   $\langle x^{-1}\rangle^{\textbf{R}}_{\pi}
   =(1/3)\langle x^{-1}\rangle_{\pi} - 1$
   of $\varphi_\text{BMS}$ (dark bars) with the upper
   and lower error-bars (grey bars), determined with nonlocal QCD sum
   rules \cite{BMS01}.
   (b) Histogram of the first nonzero Gegenbauer coefficients $a_n$
   of the BMS pion DA and the envelopes of the ``bunch'' like in (a).}
 \label{fig:histogram}
\end{figure}
where the index $i$ runs over all scalar, vector, and tensor condensates
with dim=6 \cite{BMS01,BM98};
$M^2$ is the Borel parameter,
$s_{\pi}^0$ the duality interval in the axial channel.
This sum rule allows us to determine the first ten moments
$\langle\xi^N\rangle_\pi \equiv \int_{0}^{1}
\varphi_\pi(x)(2x-1)^N dx$ of the pion DA
and independently the inverse moment
$\langle x^{-1}\rangle_{\pi}\equiv \int_{0}^{1}\varphi_\pi(x)x^{-1} dx $
quite accurately
(see in \cite{BMS01c} for an illustration).
The corresponding Gegenbauer coefficients can be determined
(see \cite{BMS01})
from this set within some error range (Fig.~\ref{fig:histogram}b)
that translates into a ``bunch'' of pion DAs
shown for $\lambda_q^2=0.4$~GeV${}^2$ in Fig.~\ref{fig:bunch}.

\begin{figure}[hbt]
 \centering{\epsfig{file=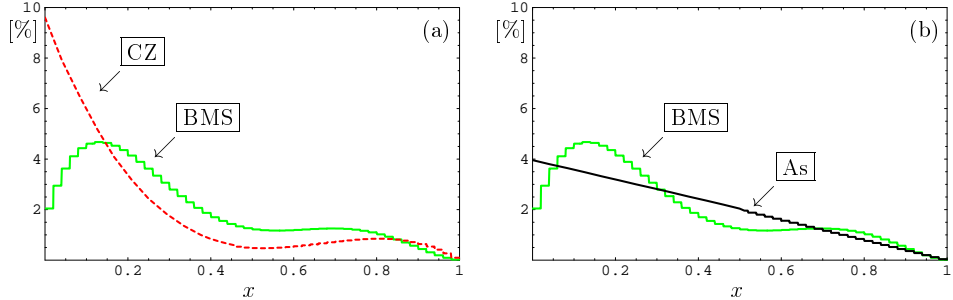,height=4cm,width=12.7cm,clip=}}
  \vspace{0.0cm}\caption{Percentage distribution (see text) of the
   first inverse moment in $x$ of the BMS model \cite{BMS01} in
   comparison with rival models.}
 \label{fig:endpoint}
\end{figure}

Their striking feature is that their endpoints are suppressed
relative to both the CZ model and the asymptotic solution.
Crudely speaking, this shape structure
is the net result of the interplay between the perturbative contribution
and the non-perturbative term $\Delta\Phi_\text{S}(x;M^2)$
(related to the scalar condensate) that dominates
the RHS of the SR in Eq.~(\ref{eq:nlcsrda}).
The fact that the function $\Delta\Phi_\text{S}(x;M^2)$
is not singular in $x$ and has a dip at the central point
of the interval $[0,1]$
is also reflected in the shapes of these DAs.
We emphasize that a suppression of the endpoint region
as strong as possible
(for a dedicated discussion we refer to \cite{SSK99})
is important in order to improve the self-consistency of perturbative QCD
in convoluting the pion DA with the specific hard-scattering amplitude
for a particular exclusive process.
In order to stress this point,
we show in Fig.~\ref{fig:endpoint}
$\langle x^{-1} \rangle_\pi$, calculated as
$\int_{x}^{x+0.02} \varphi(x) x^{-1} dx$
and normalized to $100\%$ ($y$-axis).
The main message from this figure is that $\langle x^{-1} \rangle_\pi$
receives in the endpoint region,
say between $0\leq x \leq 0.1$,
a contribution of only 17\% in the case of the BMS model,
whereas it reaches as much as 40\% for the CZ model
and still 19\% for the asymptotic solution.

\section{ANALYSIS OF THE CLEO DATA}
It was shown by Khodjamirian \cite{Kho99}
that the light-cone QCD sum-rule (LCSR) method
provides the possibility
to treat the problem of the photon long-distance interaction
(i.e., when a photon goes on mass shell)
in the $\gamma^{*}(Q^{2})\gamma(q^{2})\to\pi^{0}$ form factor
by performing all calculations for sufficiently large $q^{2}$,
using quark-hadron duality in the vector channel,
and then analytically continuing the results
to the limit $q^{2}=0$.
Schmedding and Yakovlev (SY) \cite{SchmYa99} applied these LCSRs
to the NLO of QCD perturbation theory.
More recently, we have \cite{BMS02} taken up
this sort of data processing
(i) accounting for a correct Efremov--Radyushkin--Brodsky--Lepage
 (ERBL) \cite{ER80,LB80} evolution of the pion DA
 to every measured momentum  scale,
(ii) estimating more precisely the contribution of the (next) twist-4 term,
 and
(iii) improving the error estimates in determining
 the $1\sigma$- and $2\sigma$-error contours in the $(a_2,a_4)$ plane.
Moreover, our error analysis takes into account
the variation of the twist-4 contribution
and treats the threshold effects in the running of $\alpha_s(Q^2)$
more accurately.

Our procedure is based upon LCSRs for the transition form factor
$F^{\gamma^*\gamma\pi}(Q^2,q^2 \approx 0)$ \cite{Kho99,SchmYa99}:
\begin{eqnarray}
 \label{eq:srggpi}
 F_\text{LCSR}^{\gamma^*\gamma\pi}(Q^2)
 = \frac1\pi\,\int\limits_0^{s_0}\!\!\frac{ds}{m_\rho^2}\,
    \rho(Q^2,s;\mu^2)
     e^{\left(m_\rho^2-s\right)/M^2}+
    \frac1\pi\,\int\limits_{s_0}^\infty\!\!\frac{ds}{s}\,
    \rho(Q^2,s;\mu^2) \; ,
\end{eqnarray}
following from a dispersion relation with $M^2\approx0.7$~GeV$^2$,
where $m_\rho$ is the $\rho$-meson mass and $s_0=1.5$~GeV${}^2$
denotes the effective threshold in the $\rho$-meson channel.
The spectral density
$\rho(Q^2,s;\mu^2)\equiv$
$\mathbf{Im}\left[F_\text{QCD}^{\gamma^*\gamma^*\pi}(Q^2,q^2=-s;\mu^2)\right]$
is calculated by virtue of the factorization theorem for the form factor
at Euclidean photon virtualities
$q^2_1=-Q^2 < 0$, $q^2_2= -q^2 \leq 0$ \cite{ER80,LB80,DaCh81},
and the factorization scale $\mu^2$ is fixed by SY
at $\mu^2=\mu^2_\text{SY}=5.76~\text{GeV}^{2}$.
Moreover, $F_\text{QCD}^{\gamma^*\gamma^*\pi}(Q^2,q^2;\mu^2)$
contains a twist-4 contribution,
which is proportional to the coupling $\delta^2(\mu^2)$,
defined by \cite{Kho99,NSVVZ84}
$\langle \pi (p)|g_s \bar{d}
 \tilde{G}_{\alpha\mu} \gamma^{\alpha} u|0\rangle
  =i \delta^2f_{\pi} p_\mu$,
where
$\tilde{G}_{\alpha\mu}=(1/2)\varepsilon_{\alpha\mu\rho\sigma}G^{\rho\sigma}$
and $G_{\rho\sigma}=G_{\rho\sigma}^a \lambda^a/2$.

This contribution for the asymptotic twist-4 DAs of the pion
as well as explicit expressions for the spectral density
$\rho(Q^2,s;\mu^2)$ in LO have been obtained in \cite{Kho99}
to which we refer for details.
The spectral density of the twist-2 part in NLO has been calculated
in \cite{SchmYa99}---see Eqs.\ (18) and (19) there.
All needed expressions for the evaluation of Eq.\ (\ref{eq:srggpi})
are collected in the Appendix E of \cite{BMS02},
cf.\ Eqs.\ (E.1)--(E.3).
We set $\mu^2=Q^2$ in
$F_\text{QCD}^{\gamma^*\gamma^*\pi}(Q^2,q^2;\mu^2)$
and use the complete 2-loop expression for the form factor,
absorbing the logarithms into the coupling constant
and the pion DA evolution at the NLO level \cite{BMS02}
so that
$\alpha_s(\mu^2)\stackrel{\text{RG}}{\longrightarrow}\alpha_s(Q^2)$
and
$ \varphi_\pi(x;\mu^2)
  \stackrel{\text{ERBL}}{\longrightarrow}
  \varphi_\pi(x; Q^2)=U(\mu^2 \to Q^2)\varphi_\pi(x; \mu^2)$
(RG denotes the renormalization group).
Then, we use the spectral density \hbox{$\rho(Q^2,s,Q^2)$},
derived in \cite{SchmYa99} at \hbox{$\mu^2=\mu^2_\text{SY}$},
in Eq.\ (\ref{eq:srggpi})
to obtain \hbox{$F^{\gamma^*\gamma\pi}(Q^2)$}
and fit the CLEO data over the probed momentum range,
denoted by \hbox{$\{Q^2_\text{exp}\}$}.
In our recent analysis \cite{BMS02} the evolution
$\varphi_\pi(x; Q^2)=U(\mu^2_\text{SY}\to Q^2)\varphi_\pi(x; \mu^2_\text{SY})$
was performed
\textit{for every individual point} $Q^2_\text{exp}$,
with the aim to return to the normalization scale $\mu^2_\text{SY}$
and
to extract the DA parameters $(a_2,~a_4)$
at this reference scale
for the sake of comparison with the previous SY results \cite{SchmYa99}.
In effect, for every measurement,
$\{Q_\text{exp}^2,F^{\gamma^*\gamma\pi}(Q_\text{exp}^2)\}$,
its own factorization and renormalization scheme was used
so that the NLO radiative corrections
were taken into account in a complete way.

\begin{figure}[t]
\centering{\epsfig{file=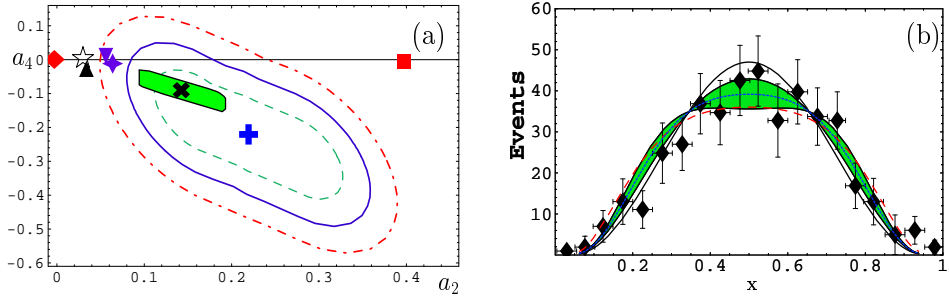,height=4cm,width=12.7cm,clip=}}
 \vspace{0.0cm} \caption{(a) Analysis of the CLEO data on
  $F_{\pi\gamma^{*}\gamma}(Q^2)$ in terms of error regions
  around the best-fit point (\ding{58})
  (broken line: $1\sigma$; solid line: $2\sigma$; dashed-dotted line: $3\sigma$)
  in the ($a_2$,$a_4$) plane
  contrasted with various theoretical models explained in the text.
  The slanted shaded rectangle represents the constraints on ($a_2,~a_4$) plane
  posed by the nonlocal QCD sum rules \protect\cite{BMS01}
  for the value $\lambda^2_q=0.4$~GeV$^{2}$.
  All constraints are evaluated at $\mu^2_\text{SY}=5.76~\text{GeV}^2$
  after NLO ERBL evolution.
  (b) Comparison of $\varphi^\text{asy}$ (solid line),
   $\varphi^\text{CZ}$ (dashed line), and the
   BMS ``bunch'' of pion DAs (strip, \protect\cite{BMS02})
   with the E791 data \protect\cite{E79102}.
   The corresponding $\chi^2$ values are:
   12.56---asy; 14.15---CZ; 10.96---BMS.}
 \label{fig:ellipses}
\end{figure}

The accuracy of this procedure is still limited
because of the uncertainties entailed by the twist-4 scale
parameter \cite{BMS02}, $k\cdot \delta^2$, with the factor $k$
expressing the deviation of the twist-4 DAs from their asymptotic
shapes (another source of uncertainty originates from the unknown
NNLO $\alpha_s$-corrections, see \cite{BMS02}).
Based on our experience with the twist-2 case, we set $k = 1 \pm 0.1$.
As a result, the final (rather conservative) accuracy estimate
for the twist-4 scale parameter
can be expressed in terms of
\hbox{$k\cdot\delta^2(1 ) = 0.19 \pm 0.04~\text{GeV}^2$} \cite{BMS02}.
To produce the complete $2\sigma$- and $1\sigma$-contours,
corresponding to these uncertainties,
we need to unite a number of regions,
resulting from the processing of the CLEO data at different values
of the scale parameter $k\cdot\delta^2$
within this admissible range \cite{BMS02}.
The obtained results for the asymptotic DA (\ding{117}),
the BMS model (\ding{54}) \cite{BMS01},
the CZ DA ({\footnotesize \ding{110}}),
the SY best-fit point ({\footnotesize\ding{108}}) \cite{SchmYa99},
a recent transverse lattice result ({\footnotesize\ding{116}})
(fourth reference in \cite{Lattice}),
and two instanton-based models,
viz., (\ding{72}) \cite{PPRWK99} and (\ding{70})
(using in this latter case $m_q=325$~MeV, $n=2$, and $\Lambda=1$~GeV) \cite{PR01},
are displayed in Fig.~\ref{fig:ellipses}(a)
varying the twist-4 scale parameter $k\cdot \delta^2$
in the interval
$[0.15 \leq k \cdot \delta^2 \leq 0.23]~\text{GeV}^2$.
The important points to observe from this figure are these:
(i) the nonlocal QCD sum-rule constraints,
    encoded in the slanted shaded rectangle,
    are in rather good overall agreement with the CLEO data
    at the $1\sigma$-level;
(ii) the CZ model and the asymptotic DA are ruled out at least
     at the $3\sigma$-level.
These findings are not significantly changed,
even if one allows an extreme twist-4 uncertainty of 30\%,
or if excluding the low-momentum-transfer data tail---say,
up to $Q^2_\text{exp}= 3$~GeV${}^{2}$ \cite{BMS02}.
In the first case, the asymptotic DA is outside the 3$\sigma$-error ellipse,
whereas in the second case it remains outside the 2$\sigma$ region
with the instanton-inspired models
just at the 2$\sigma$-ellipse boundary and the CZ model always far outside.

\section{E791 DATA: CONSTRAINTS FROM DIFFRACTIVE DI-JETS PRODUCTION}
An independent source of experimental data
to constraint the shape of the pion DA
is provided by the E791 Fermilab experiment \cite{E79102}.
Unfortunately, these data are affected by inherent uncertainties
and their theoretical explanation by different groups
\cite{Che01,NSS01,BISS02}
is still controversial so that they cannot be used
to exclude some optional model.
For our exposition here the important point
is to show that our predictions for this process
are not conflicting the E791 data using for all considered models
the \textit{same} calculational framework,
notably the convolution approach of \cite{BISS02}.
The results of the calculation are displayed in Fig.\ \ref{fig:ellipses}(b)
making evident that the E791 data are relatively in good agreement
with our prediction---especially, in the middle $x$ region,
where our DAs ``bunch'' has the largest uncertainties (see Fig.~\ref{fig:bunch}a).
Note, however, that all theoretical predictions
shown in this figure
are not corrected for the detector acceptance.
For a more precise comparison, this distortion must be taken into
account.

\section{CONCLUSION}
Both analyzed experimental data sets
(CLEO \cite{CLEO98} and Fermilab E791 \cite{E79102})
converge to the conclusion that the pion DA is not everywhere
a convex function, like the asymptotic one,
but has instead two maxima
with the end points ($x=0,1$) strongly suppressed---in contrast to the CZ DA.
These two key dynamical features of the DA are both controlled
by the QCD vacuum inverse correlation length $\lambda_q$,
whose value suggested by the CLEO data analysis
is $\lambda_q^2 \sim 0.4$~GeV${}^2$ in good compliance
with the QCD sum-rule estimates and lattice computations.

\section*{ACKNOWLEDGEMENTS}
This work was supported in part by INTAS-CALL 2000 N 587,
the RFBR (grants 03-02-16816, 03-02-04022-NNIO and 03-02-26737),
the Heisenberg--Landau Programme,
the COSY Forschungsprojekt J\"ulich/Bochum,
and the Deutsche Forschungsgemeinschaft (DFG).
A.\ P.\ B.\ would like to thank the conference organizers
for their hospitality and support.


\begin{thebibliography}{99}
\bibitem{CZ84} V.L.~Chernyak and A.R.~Zhitnitsky,
               Phys.\ Rep.\ \textbf{112} (1984) 173.
\bibitem {Ste99} N.G.~Stefanis,
                 Eur.\ Phys.\ J.\ directC \textbf{1}, 7 (1999).
\bibitem{MR86} S.V.~Mikhailov and A.V.~Radyushkin,
               JETP Lett.\ \textbf{43} (1986) 712;
               Sov.\ J.\ Nucl.\ Phys.\ \textbf{49} (1989) 494;
               Phys.\ Rev.\ D \textbf{45} (1992) 1754;
               A.P.~Bakulev and A.V.~Radyushkin,
               Phys.\ Lett.\ B \textbf{271} (1991) 223;
               A.P.~Bakulev and S.V.~Mikhailov,
               Z.\ Phys.\ C \textbf{68} (1995) 451.
\bibitem{ER80} A.V.~Efremov and A.V.~Radyushkin,
                 Phys.\ Lett.\ B \textbf{94} (1980) 245;
                 Theor.\ Math.\ Phys.\ \textbf{42} (1980) 97.
\bibitem{LB80} G.P.~Lepage and S.J.~Brodsky,
                 Phys.\ Lett.\ B \textbf{87} (1979) 359;
                 Phys.\ Rev.\ D \textbf{22} (1980) 2157.
\bibitem{BF89} V.M.~Braun and I.E.~Filyanov,
               Z.\ Phys.\ C \textbf{44} (1989) 157.
\bibitem{BJ97} V.M.~Belyaev and M.B.~Johnson,
               Phys.\ Rev.\ D \textbf{56} (1997) 1481;
               Phys.\ Lett.\ B \textbf{423} (1998) 379.
\bibitem{BM98} A.P.~Bakulev and S.V.~Mikhailov,
               Phys.\ Lett.\ B 436 (1998) 351.

\bibitem{BMS01} A.P. Bakulev, S.V. Mikhailov, and N.G. Stefanis,
                 Phys.\ Lett.\ B \textbf{508} (2001) 279.
\bibitem{PPRWK99} V.Y.~Petrov et al.,
                 Phys.\ Rev.\ D \textbf{59} (1999) 114018.
\bibitem{PR01} M.~Praszalowicz and A.~Rostworowski,
                 Phys.\ Rev.\ D \textbf{64} (2001) 074003;
                 ibid. \textbf{66} (2002) 054002.
\bibitem{ADT00} I.V.~Anikin, A.E.~Dorokhov, L.~Tomio,
                 Phys.\ Lett.\ B \textbf{475} (2000) 361.
\bibitem{Lattice} G.~Martinelli and C.T.~Sachrajda,
                 Phys.\ Lett.\ B \textbf{190} (1987) 151;
                  D.~Daniel et al.,
                 Phys.\ Rev.\ D \textbf{43} (1991) 3715;
                  M.~Burkardt and H.~El-Khozondar,
                 Phys.\ Rev.\ D \textbf{60} (1999) 054504;
                  M.~Burkardt and S.~Seal,
                 Phys.\ Rev.\ D \textbf{65} (2002) 034501;
                  S.~Dalley and B.~van de Sande,
                 Phys.\ Rev.\ D \textbf{67} (2003) 114507;
                  L.~{\uppercase{d}el} Debbio et al.,
                 Nucl.\ Phys.\ Proc.\ Suppl.\  \textbf{83} (2000) 235;
                 \textbf{119} (2003) 416.
\bibitem{CLEO98} J.~Gronberg et al., CLEO Collaboration,
                 Phys.\ Rev.\ D \textbf{57} (1998) 33.
\bibitem{SchmYa99} A.~Schmedding and O.~Yakovlev,
                   Phys.\ Rev.\ D \textbf{62} (2000) 116002.
\bibitem{BMS02} A.P.~Bakulev, S.V.~Mikhailov, and N.G.~Stefanis,
                Phys.\ Rev.\ D \textbf{67} (2003) 074012;
                hep-ph/0303039. 
\bibitem{Kho99} A.~Khodjamirian,
                Eur.\ Phys.\ J.\ C \textbf{6} (1999) 477.
\bibitem{E79102} E.M.~Aitala et al., Fermilab E791 Collaboration,
                 Phys.\ Rev.\ Lett.\ \textbf{86} (2001) 4768.
\bibitem{Che01} V.~Chernyak,
                Phys.\ Lett.\ B \textbf{516} (2001) 116.
\bibitem{NSS01} N.N.~Nikolaev, W.~Sch\"afer, and G.~Schwiete,
                Phys.\ Rev.\ D \textbf{63} (2001) 014020.
\bibitem{BISS02} V.M.~Braun et al.,
                 Nucl.\ Phys.\ B \textbf{638} (2002) 111.
\bibitem{SVZ79} M.A. Shifman, A.I. Vainshtein, and V.I. Zakharov,
               Nucl.\ Phys.\ B \textbf{147} (1979) 385,
               448,
               519.
\bibitem{BM02} A.P.~Bakulev and S.V.~Mikhailov,
               Phys.\ Rev.\ D \textbf{65} (2002) 114511.
\bibitem{BMS01c} A.P.~Bakulev, S.V.~Mikhailov, and N.G.~Stefanis,
in ``Proceedings of the 36th Rencontres De Moriond On QCD And Hadronic Interactions,
  17--24 Mar 2001, Les Arcs, France'', Ed.~J.~Tran Thanh Van, Singapour,
  World Scientific, 2002, pp. 133--136 [hep-ph/0104290].
\bibitem{SSK99} N.G.~Stefanis et al.,
                Phys.\ Lett.\ B \textbf{449} (1999) 299;
                Eur.\ Phys.\ J.\ C \textbf{18} (2000) 137.
\bibitem{DaCh81} F.~del Aguila and M.K.~Chase,
                 Nucl.\ Phys.\ B \textbf{193} (1981) 517;
                 E.P.~Kadantseva, S.V.~Mikhailov, and A.V.~Radyushkin,
                Sov.\ J.\ Nucl.\ Phys.\ \textbf{44} (1986) 326.
\bibitem{NSVVZ84} V.A.~Novikov et al.,
                  Nucl.\ Phys.\ B \textbf{237} (1984) 525.
\end{thebibliography}
\end{document}